\documentstyle{article}
\begin{document}
\begin{large}

\centerline{\Large\bf Optical spectra measured on cleaved surfaces of}

\centerline{\Large\bf double-exchange ferromagnet La$_{1-x}$Sr$_x$MnO$_3$}

\bigskip
\centerline
{
\underline{Koshi Takenaka},$^1$ Kenji Iida, Yuko Sawaki, Shunji Sugai
}

\centerline
{
Yutaka Moritomo,$^*$ and Arao Nakamura$^*$
}

\bigskip

\centerline{Department of Physics, Nagoya University, Chikusa-ku, Nagoya 464-8602, Japan}

\centerline{$^*$CIRSE, Nagoya University, Chikusa-ku, Nagoya 464-8603, Japan}

\vspace{2cm}

$^1$author to whom correspondence should be addressed

\hspace{1cm}institution: Department of Physics, Faculty of Science, Nagoya University

\hspace{1cm}address: Furo-cho, Chikusa-ku, Nagoya 464-8602, Japan

\hspace{1cm}E-mail: k46291a@nucc.cc.nagoya-u.ac.jp

\hspace{1cm}Fax: +81-52-789-2933

\vfill

\centerline{\bf Abstract}

\begin{quote}
Optical reflectivity spectra were measured on cleaved surfaces of
La$_{1-x}$Sr$_x$MnO$_3$ single crystals ($x$$=$0.175, 
$T_{\rm C}$$=$283 K) over a temperature range 10$-$295 K. The
optical conductivity $\sigma(\omega)$ shows, keeping single-component
nature, incoherent-to-coherent crossover with increase of electrical
conductivity. The $\sigma(\omega)$ spectrum of low-temperature
ferromagnetic-metallic phase (10 K) exhibits a pronounced Drude-like
component with large spectral weight, contrary to the previous
result. The present result indicates that the optical spectrum of the
manganites is sensitive to condition of sample surfaces.
\end{quote}

PACS numbers: 78.30.$-$j, 71.27.+a, 71.30.+h

Substance Classification: S1.2, S10.15

\newpage

Optical reflectivity studies are important from both fundamental and
practical viewpoints, since it enables us not only to deduce the
dielectric function, but also to examine separately two key-elements
of the charge transport, {\it i.e.,} the carrier density (or Drude
weight) and the scattering time. The charge transport is one of the
central concerns for both camps. However, for the case of the
double-exchange ferromagnetic-metal manganites, which have recently
attracted renewed interest because of its intriguing phenomenon, 
colossal magnetoresistance (CMR) \cite{cmr}, the previous
results of the reflectivity studies are rather confused and
controversial \cite{okimoto97,kim98,boris99}.

We report the optical reflectivity spectra $R(\omega)$ of
La$_{1-x}$Sr$_x$MnO$_3$ ($x$$=$0.175, $T_{\rm C}$$=$283 K) measured
on {\it cleaved} surfaces of single crystals over a wide temperature
range (10$-$295 K). The optical conductivity $\sigma(\omega)$
exhibits a single-component with large spectral weight, contrary to
the previous results. The present result indicates that the optical
spectrum of the manganites is very sensitive to condition of
surfaces, which can partly explain the above confusion
\cite{takenaka99}.

Single crystals of La$_{1-x}$Sr$_x$MnO$_3$ were grown by a floating
zone method \cite{urushibara95}. The size of the cleaved surface was
at largest 1.0$\times$1.0 mm$^2$. Temperature-dependent optical
reflectivity was measured using a Fourier-type interferometer
(0.02$-$1.6 eV) and a grating spectrometer (0.8$-$6.6 eV). The
experimental error of reflectivity, $\Delta R$, determined by the
reproducibility, is less than 1\% for the far-IR to visible region
and less than 2\% for the ultraviolet region.

Figure 1 shows the temperature-dependent (10$-$295 K) reflectivity
spectra measured on the cleaved surfaces of
La$_{0.825}$Sr$_{0.175}$MnO$_3$ single crystals on a logarithmic
energy scale in the range 0.01$-$6.0 eV. The Curie temperature
$T_{\rm C}$ measured by dc resistivity was 283 K. With decreasing
temperature, the reflectivity spectrum $R(\omega)$ changes gradually
from insulating to metallic behavior: the reflectivity edge at about
1.6 eV becomes sharpened, though its position does not shift
appreciably, and the optical phonons are screened corresponding to
increase of electrical conductivity. Below 100 K, the optical phonons
almost fade away and the spectrum is characterized by a sharp edge
and a large spectral weight below it. The present spectrum measured
on cleaved surfaces is much higher at a mid-IR-to-visible region
compared with the previous results measured on polished surfaces
\cite{okimoto97,kim98}.

In order to make more detailed discussions, we deduce optical
conductivity $\sigma(\omega)$ (Fig. 2) from $R(\omega)$ shown in
Fig. 1 via a Kramers-Kronig transformation. We measured reflectivity
spectra at each temperature below 6.6 eV and above 6.6 eV we assumed
the data measured at room temperature (295 K) using a Seya-Namioka
type spectrometer for vacuum-ultraviolet (VUV) synchrotron radiation
up to 40 eV at Institute for Molecular Science, Okazaki National
Research Institutes. Such a procedure is possible and reasonable
because the variation of $R(\omega)$ at 6.0$-$6.6 eV is negligible
small and the data below 6.6 eV could be connected smoothly with the
VUV-data. For the extrapolation at the lower-energy part, we assumed
a constant $R(\omega)$ for the insulating phase ($T$$=$295 K). For
the metallic phase ($T$$\le$278 K), we make a smooth extrapolation
using a Hagen-Rubens formula. Extrapolating parameter $\sigma(0)$
is roughly in accord with the dc value \cite{okuda98}. Variation
of the extrapolation procedures had negligible effect on
$\sigma(\omega)$ in the energy region of interest (0.03$-$6.0 eV).

The optical conductivity $\sigma(\omega)$ shows, keeping
single-component nature, incoherent-to-coherent crossover with
decreasing temperature: Above $T_{\rm C}$, the $\sigma(\omega)$
spectrum is characterized solely by a broad peak centered at $\sim$1.5
eV. At the temperature range 278$-$220 K, this broad peak gradually
develops and its position shifts downwards as $T$ decreases but a
Drude-like component is {\it not} confirmed in the present
experiment, though the material is ferromagnetic-metallic. Below 155
K, on the other and, the spectrum exhibits a single Drude-like
component centered at $\omega$$=$0 and it becomes narrow as $T$
decreases without increase of spectral weight.

Integrated spectral weight defined as

$$N^*_{\rm eff}(\omega)={{2m_0V} \over {\pi e^2}} \int_0^\omega
\sigma(\omega ')d\omega '\eqno(1) $$

\noindent
($m_0$: a bare electron mass; $V$: the unit-cell volume) represents
an effective density of carriers contributing to optical transitions
below a certain cutoff energy $\hbar\omega$ (inset of Fig. 2). The
characteristic single-component which shows incoherent-to-coherent
crossover consists of the spectral weight transferred from the two
bands at $\sim$3 eV and at $\sim$5 eV. The present result suggests
that the exchange-split down-spin band \cite{furukawa95} consists of
two bands. Because the curves of $N^*_{\rm eff}(\omega)$ merge into
almost a single line above 6 eV, the down-spin band does not seem to
split to more than two bands. Imperfect convergence is most likely due
to the increasing experimental error on $\sigma(\omega)$ with
$\omega$. This partly justifies our procedure that the data above 6.6
eV at room temperature is connected with the data measured at each
temperature.

Finally we show that the discrepancy between our result and the
previous result may originate from sensitivity of the spectrum to
the condition of the surface. In Fig. 3 are shown the
room-temperature (295 K) reflectivity spectra measured on a cleaved
surface (solid line) as well as that measured on a surface polished by
lapping films with diamond powder (dotted line). It is found that
polishing dramatically alters $R(\omega)$ for the ferromagnetic metal
La$_{0.70}$Sr$_{0.30}$MnO$_3$ [Fig. 3(b)] whereas it affects only
slightly the spectrum for the undoped LaMnO$_3$ [Fig. 3(a)]. The
previous data \cite{okimoto97,kim98} resembles closely the spectrum
measured on the polished surface. The damage of the surface probably
localizes the carriers. However, light with long wavelength reaches
inside the bulk and hence $R(\omega)$ recovers the intrinsic spectrum, 
which is consistent with the observation that the discrepancy almost
disappears below 0.03 eV \cite{sd}. \lq\lq Small Drude weight" may
originate from the above restoration process [inset of Fig. 3(b)].

In summary, we have reported the temperature-dependent optical spectra
of the prototypical double-exchange system La$_{1-x}$Sr$_x$MnO$_3$
measured on the cleaved surfaces. The optical conductivity spectra are
characterized by a single component which shows incoherent-to-coherent
crossover with increase of the electrical conductivity and a
pronounced Drude-like component is observed for the
ferromagnetic-metallic phase at low temperatures. The present result
indicates that the optical spectrum of the doped manganites is
sensitive to condition of surfaces. The charge dynamics of the doped
manganites might be extremely sensitive to static imperfections.

We would like to thank M.~Kamada, M.~Hasumoto, and R.~Yamamoto for
the help in the experiments. One of us (K.T.) is also grateful to
N.~E.~Hussey for a critical reading of the manuscript. This work was
partly supported by Grant-in-Aid for Scientific Research from the
Ministry of Education, Science, and Culture of Japan and by CREST of
JST.

\newpage
\noindent{\Large\bf Figure Captions}

\bigskip

\noindent
Fig. 1  Temperature-dependent optical reflectivity spectra measured on
cleaved surfaces of La$_{0.825}$Sr$_{0.175}$MnO$_3$ ($T_{\rm C}$$=$283 K).

\bigskip

\noindent
Fig. 2  Temperature-dependent optical conductivity spectra of
La$_{0.825}$Sr$_{0.175}$MnO$_3$ deduced from the reflectivity spectra
measured on the cleaved surfaces (shown in Fig. 1) via a Kramers-Kronig
transformation. Inset: Effective carrier number per Mn-atom
$N^*_{\rm eff}(\omega)$ defined as the integration of $\sigma(\omega)$.

\bigskip

\noindent
Fig. 3  Room-temperature (295 K) optical reflectivity spectra measured
on the cleaved (solid line) and polished (dotted line) surfaces: (a)
LaMnO$_3$ and (b) La$_{0.70}$Sr$_{0.30}$MnO$_3$. Inset: Optical
conductivity spectra deduced from the Kramers-Kronig transformation of
the reflectivity spectra shown in Fig. 3(b).

\end{large}

\begin{thebibliography}{99}
\bibitem{sd} We estimated skin-depth as the inverse of absorption
  coefficient $\alpha(\omega)$$=$$2\omega k(\omega)/c$ and confirmed
  that the skin-depth becomes longer below 0.1 eV for $x$$=$0.30.
\end{thebibliography}
\end{document}